# NEG pumps: Sorption mechanisms and applications

*P. Manini, E. Maccallini*

SAES GETTERS S.p.A., viale Italia 77, 20020, Lainate, Milan, Italy

**Abstract**
This paper reports the main physical and chemical properties of NEG materials, sorption mechanisms and use of NEG pumps from high to extreme high vacuums.

**Keywords**
NEG, NEG pumps; getter; gas sorption; sintering; high vacuum; UHV; XHV; zirconium; Vanadium.

## 1  Introduction: fundamentals of Bulk Getters/NEG

### 1.1  Requirements on NEG pumps and key features

Getter materials in bulk getter pumps bind gas by chemical reaction. Thus, they must be chemically reactive towards residual gases typically occurring in vacua (CO, $CO_2$, $N_2$, $O_2$, $H_2O$, $H_2$, etc.). Additionally, however, they should provide easy handling in contact with atmospheric air when being mounted.

Getters are generally classified into two distinctive families, i.e. Non-Evaporable Getters (NEG) and Evaporable Getters (EG).

NEG are made of reactive metals, like titanium, zirconium, vanadium and their alloys, and are typically formed as compressed or sintered powders. Due to their chemical reactivity, the getter powder surface is easily passivated with some monolayers of oxides and carbides during the manufacturing process. This avoids further reaction with atmospheric gases and protect the material during handling in air. Once mounted under vacuum, the passivation layer must be removed to free the underlying reactive surface and to start the chemisorption process. This is accomplished by a thermal treatment, called "activation" which promotes the diffusion of the chemically bond oxygen, carbon and nitrogen atoms from the surface into the getter bulk. The efficiency of the diffusion process strongly depends on the getter composition and microstructure. Due to this, activation temperature and duration can significantly change from one getter alloy to another.

Also, Evaporable Getters are made of relatively stable alloys which includes a reactive metal, such as Ba, Ca or Ti. They can be handled in air during mounting and assembling, however, differently from NEG, after being inserted in the vacuum device the reactive metal is evaporated. The evaporation process is carried out for a very short period of time (seconds) at relatively higher temperatures (≈900°C) to generate a large surface area upon which molecules are chemisorbed. For example, the common barium getters are used to generate elemental barium through a reaction of a mixture of a $BaAl_4$ alloy and nickel at $800°C$ to $1250°C$ under high vacuum. During the heating process, Al reacts with Nickel, freeing barium atoms which evaporate from the getter container and condense as a reactive layer on the opposing surface.

Non-Evaporable Getters are used primarily in applications where either evaporation of metals under vacuum is undesired or a surface for depositing a metal film is unavailable. A key distinctive feature of NEG is their large pumping speed per unit volume. As a direct consequence of this feature, NEG pumps typically have a compact package and can be mounted in small footprint or space-limited systems. NEG pumps are also particularly efficient in pumping hydrogen, the main residual gas in typical UHV systems. Additional features are the small weight, the absence of vibration (no moving parts), marginal power consumption, no maintenance and negligible interference with magnetic field (magnetic permeability of mostly used getter is <1,001).

## 1.2 Activating NEG pumps

Getters bind gases at the surface. Thus, large available surfaces are desired. After being placed in a vacuum, such surfaces, as any other surface, require decontamination from physically bound gases. This is done by baking.

The oxide/nitride layer on the passivating surface has to be dissolved in order to allow reactions between the getter material and gases. This step involves further heating under vacuum. In contrast to physically bound gases, the chemical bonds between NEG and oxygen, carbon and nitrogen atoms are too strong to be separated by heating. Therefore, even at temperatures around 1000 K, equilibrium pressures of, e.g., oxygen and nitrogen, above their corresponding compounds are in the range of only $10^{-15}$ Pa [1].

At the same time, high temperature increases the diffusion rates of oxygen, carbon and nitrogen ions in the getter. Following the concentration gradient, the ions migrate into the bulk, the surface returns to metal state, and thus, getter regains the ability to bind gases. This diffusion mechanism has been clearly showed in a number of studies correlating the surface concentration of carbon and oxygen with the activation treatment [2-3]. The temperature and time required for the diffusion process to take place depends on the type of getter material. Typical results are illustrated in Fig. 1 for two different alloys. The former (St 707®) is a Zr-V-Fe alloy fully activated at 500°C, the latter (St 101®) a Zr-Al alloy -activated at 800°C. Such activation process cause in both cases the diffusion of carbon and oxygen in the bulk and the emergence of metallic zirconium. In the activation process, the activation temperature is the most important parameter. However, as the atoms diffusion depends on the square root of time, time also plays a role, even though minor. Increasing the activation time can be applied when it is not possible to increase the temperature beyond a certain limiting value. The maximum timeframe is usually predefined by the application. It is generally comprised between one to a few hours, even though it can be much shorter (e.g. ten seconds, as in the lamp industry) or even one day long (e.g. in large vacuum vessels).

In certain cases, partial activation of the getter surface is sufficient (see Fig. 2). It has to be considered in fact that the activation efficiency is directly proportional to the initial nominal pumping speed. A 60% activation efficiency, for example, simply means that the NEG pump initial speed is 60% of the nominal value, which might be enough or acceptable in a given application. As previously mentioned, gases are released by the getter during the activation process. Physisorbed gases are first given off when temperature raises from room temperature to ≈200°C. Increasing the temperature, hydrogen and methane are then released. Hydrogen ions, naturally present as a solid solution in the getter volume, diffuse to the surface where they recombine in hydrogen molecules and are desorbed. Methane (and other hydrocarbons) can be generated on the getter surface by the thermally activated recombination of carbon and hydrogen atoms. After a suitable conditioning, desorption rate of physisorbed gases and methane is significantly reduced. Hydrogen remains the main gas which is released during the further temperature ramping up. Even though part of the emitted hydrogen is removed by the pumping group, pressure as high as $10^{-4}$-$10^{-5}$ mbar are not uncommon during the activation, as the getter volume acts as a reservoir for hydrogen emission. At the end of the activation process, during NEG pump cooling down, hydrogen in the gas phase is finally fully re-adsorbed by the getter. These steps are clearly shown in Fig. 3 which provides the type and amount of different gases released during the conditioning and the activation steps for a St 172® based getter pump. Even though the activation process is necessary to turn the passivated layer into a metallic active surface, the emission of physisorbed gases and hydrogen should be possibly minimized as it can be a source of contamination or be a burden for the vacuum systems. In fact, a large gas emission can result in a very long and tedious activation process if not efficiently handled by the pumps. To this purpose NEG pumps with reduced getter mass and using very low hydrogen equilibrium isotherm alloy are in principle preferable.

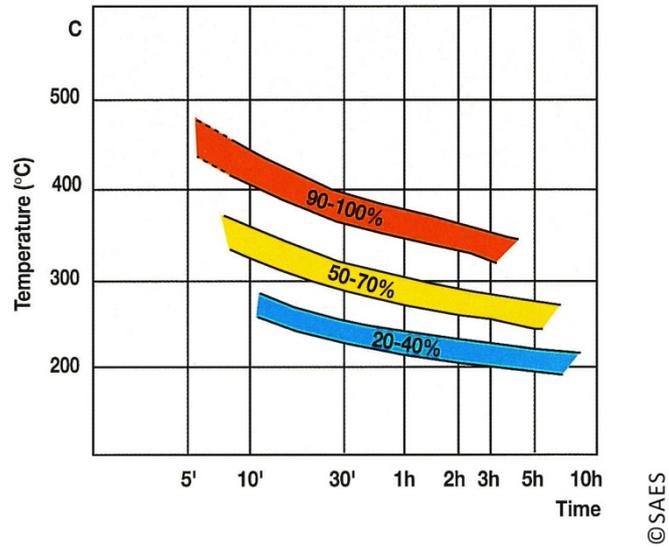

**Fig. 1:** Activation diagram for St 101® and St 707® Zr alloys (see text).  * standard activating conditions

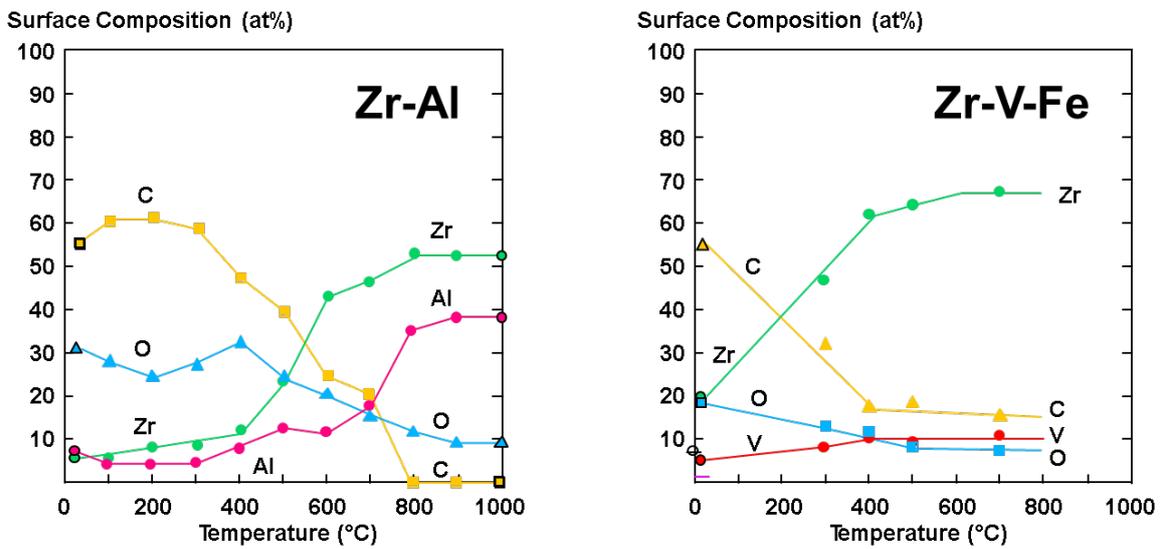

**Fig. 2:** XPS study on the modification of the surface chemistry of St 101® and St 707® Zr alloys as a function of the heat treatment (see text).

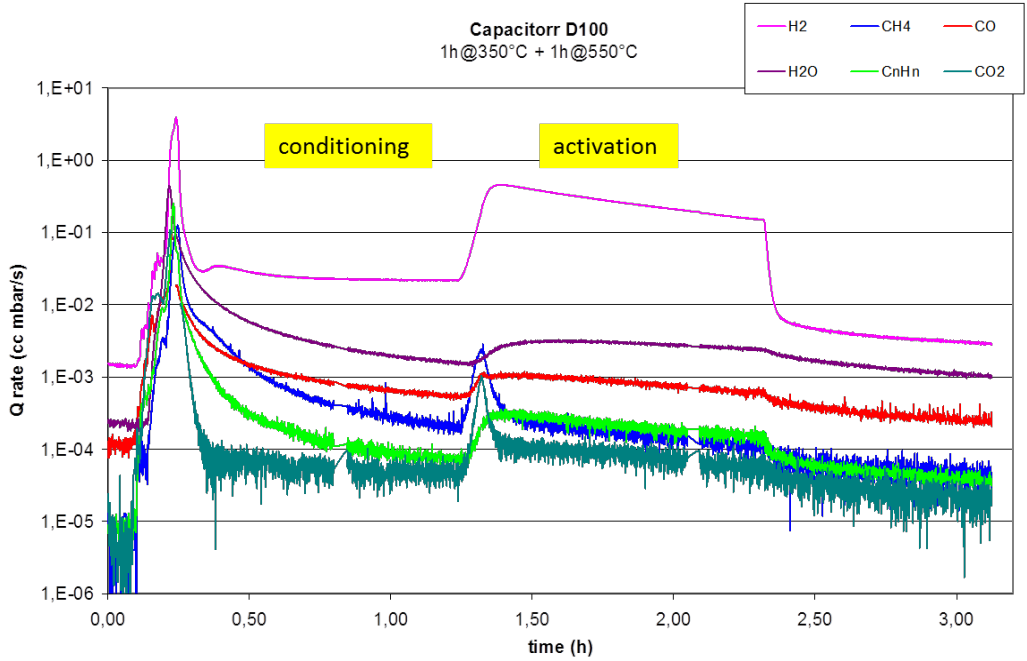

**Fig. 3:** Outgassing test on a NEG pump during conditioning and activation. Gas species have been measured by means of a quadrupole mass spectrometer.

## 1.3 Binding different gas species with bulk getters/NEG

Gas species, generally appearing in different proportions in vacuum applications, can be differentiated with respect to sorption by bulk getters:

a) Chemically active gases reacting with the getter surface.

CO, $CO_2$, $O_2$, and $N_2$ gases bind to the getter surface in several steps where they dissociate and are trapped as O, C or N atoms. This process takes place at room temperature and lower temperatures. The produced ions (carbides, oxides, and nitrides) stays on the surface and diffuse from the surface and into the bulk of the getter material only at high temperature. Bonding is irreversible. Once O, N and C are chemically bonded they cannot be released in the vacuum anymore. The same applies to $H_2O$, which dissociates at the surface and is bound as O and H.

b) Chemically active gases dissolving within the complete getter volume.

Hydrogen and its isotopes (deuterium and tritium) react in the following way. Hydrogen molecule decomposes on the getter surface into protons which then diffuse throughout the getter volume. Given the small size, protons have a very large diffusivity and can diffuse quickly inside the getter volume at or even below room temperature. The amount of hydrogen in the getter volume is in a temperature-dependent equilibrium with the outside partial pressure of hydrogen.

*Sievert's* law describes this equilibrium (see Fig. 4):

$$\log p_H = A + 2\log q_H - \frac{B}{T}, \qquad (1)$$

$p_H$: hydrogen partial pressure;  $A, B$: constants;  $q_H$: amount of hydrogen in getter.

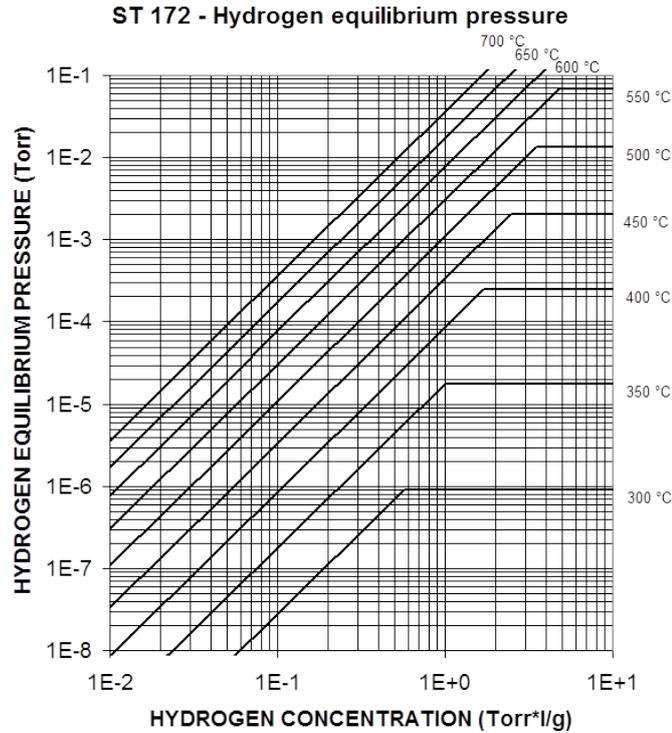

**Fig. 4:** Behavior of Zr alloy (ex. St 172) vs $H_2$ at selected temperatures according to *Sievert's* law (straight lines with slope 2). For higher $H_2$ concentrations hydrides are formed (plateau region).

Examples of A, B parameters are provided in Tab. 1 for three different alloys having the same activation temperature (pressure is given in Torr, amount in Torr·l/(g of getter), and temperature in Kelvin).

Table 1. A and B parameters in Sievert law for different NEG alloys.

| ALLOY | St 707® | St 172 | ZAO®1 |
|---|---|---|---|
| Chemical composition | Zr-V -Fe | St 707® + Zr | Zr–V-Ti-Al |
| A | 4,8 | 4,45 | 3,64 |
| B (Kelvin) | 6116 | 5730 | 7290 |

As a general rule, getter materials with low $H_2$- equilibrium pressure are preferable for high and ultra-high vacuum applications. They ensure a lower hydrogen emission during heating and a potentially larger hydrogen storage capacity at a given temperature. Sorption of hydrogen is a reversible process in the concentration range covered by the Sievert's law. This can be exploited to store and released hydrogen under controlled conditions. Once a large amount of hydrogen has been sorbed by a NEG, it is always possible to remove it by heating the getter under pumping (for example with a turbo-molecular pump). This procedure, called "regeneration", allows to extract the sorbed hydrogen and deplete the getter volume. After the regeneration the getter is ready to start a new hydrogen sorption cycle.

The time t required to remove the hydrogen from the initial value $q_i$ to the final value $q_f$ depends on getter temperature T according to Eq. (2):

$$t = \frac{M}{F}\left(\frac{1}{q_f} - \frac{1}{q_i}\right) 10^{-\left(A - \frac{B}{T}\right)} \qquad (2)$$

M is the getter mass and F is the pumping speed of the auxiliary pump (l/s).

Beyond a certain solubility value, the exceeding sorbed hydrogen is turned into metal hydrides. With more hydrogen sorbed, more hydrides are formed and a bi-phasic system (hydrogen in solid solution in equilibrium with metal hydrides) is generated, which does not obeys to Sievert's law. The onset for Metal hydrides generation depends on the alloy type and its temperature, the lower the temperature the larger the fraction of metal hydrides with respect to solid solution. This behavior is clearly showed in Fig. 4 where the equilibrium isotherm for St 172 are plotted vs equilibrium temperature and hydrogen concentration.

The straight lines with slope 2 are Sievert's law isotherms, while the flat plateau corresponds to the bi-phasic zone, where hydrogen in solid solution and metal hydrides are in equilibrium. From a practical point of view one can take advantage of this behavior to keep in the plateau region a very stable and controlled hydrogen equilibrium pressure by simply adjusting the temperature.

Sorption of a large amount of hydrogen changes the alloy lattice parameters and introduces macroscopic mechanical stresses. These can finally develop in cracks and defect lines which undermine the solidity and the mechanical integrity of the getter element (sintered disks or compressed pellet). For traditional getters like St 707® and St 172 the concentration (embrittlement) limit is approximately set at 25hPa·l/g. If the getter pump is supposed to be thermally cycled (activated/regenerated) many times it is even preferable to use a lower limit, such as 15hPa·l/g as thermal fatigue in hydride getters is also a source of mechanical stresses. To overcome this issue, getters with higher embrittlement limit have been developed which can handle larger hydrogen load see Section 6). This is particularly interesting in high vacuum application where NEG pumps have to cope with larger hydrogen flow rate and partial pressures. One of these getters is the quaternary alloy ZAO®1 (see Section 6).

c) Chemically passive gases not reacting with the getter material or reacting only under high-temperature conditions.

Simple hydrocarbons are relatively inactive towards NEG materials. This is particularly true for methane, $CH_4$. They react at higher temperatures if the chemical bonds between carbon and hydrogen crack near hot surfaces. Binding then occurs and produces $H_2$ or carbides. Long-chain hydrocarbons also physisorb to the getter surface. Getters do not sorb noble gases because no chemical reaction is possible. This is exploited to purify noble/inert gases with getters.

NOTE: The difference between a) and b) sorption mechanisms determines two different thermal processes to restore the NEG pumping characteristics: the reactivation and the regeneration processes respectively. In the first case, active gases will cover and saturate the NEG surface. For this reason, pumping speed will be reduced to zero with full NEG surface coverage. The reactivation will restore the pumping speed for all gases. While by regeneration process, the capacity for $H_2$ will be reversibly restored.

Typically, the reactivation process lasts 1-3 hours. While the regeneration process lasts hours/days depending on the parameter values included in Eq. (2).

## 1.4 Composition of bulk getters/NEG

Following the list of required properties mentioned above, metals are suitable, particularly group 4 elements of the periodic table (titanium, zirconium, hafnium), and also, tantalum, niobium, and thorium. In practice, apart from titanium, typical materials are specially developed alloys of zirconium.

A more recent improvement in alloy formulation is the Zr-based alloy commercially available under the trademark ZAO® 1 (Zr –V-Ti), which is characterized by significantly larger gas capacity for active gases, hydrogen included, low gas desorption during activation and negligible dust emission (see Section 6). This alloy can either be used in High Vacuum and in UHV applications.

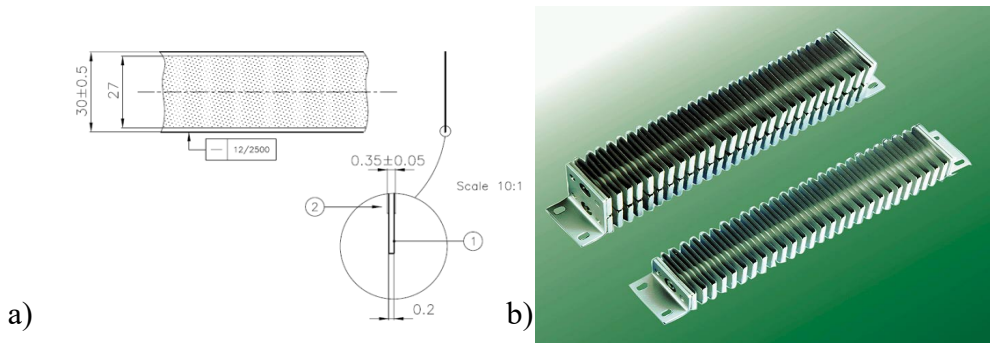

a)   b)

**Fig: 5:** Strip getter. (a) Standard strip getter 30D, dimensions given in millimeters; (b) strip getter folded to meander shape with fixing device (getter module). Courtesy of *SAES Getters S.p.A.*

## 1.5    Bulk getter/NEG stock

Getter alloys are commonly produced by melting reactive metals under vacuum. The metallic ingot is crushed into pieces and granules which are then milled into powder having suitable granulometry. The getter powder is generally consolidated and shaped in different configurations such as compressed getter strips, pellets or rings:

- Getter strips (see Fig. 5)

    Here, the getter material is rolled on a carrier tape from one or both sides using high force. The carrier is made of nickel-plated iron sheet or constantan (nonmagnetic). Thickness of the getter layer is around 100 microns.

- Getter pellets

    If the getter material is not too brittle, it can be compressed to pellets.

- Getter rings

    Getter material is pressed into metal rings with U-shaped cross section. The shape of the ring improves activating when heated with high-frequency excitation.

The powders can also be cold laminated upon suitable substrates to provide a relatively flexible getter strip.

The compression and lamination technology of getter powders is a relatively simple and well assessed technology which provides mechanically stable items. It presents however two drawbacks:

- reduction of the exposed getter surface and the open accessible pores
- generation of dust (which might be unwanted in some applications).

A possible way to overcome these two limitations has been developed in the 80s consisting in an effective consolidation of getter powders achieved by vacuum sintering. Getter alloys are sintered at high temperature under vacuum to highly porous, self-supporting bodies. During sintering, adjacent getter grains bond together by surface melting, creating a single network of particle not prone to release dust. The process can be optimized in order to keep the void fraction and the available getter surface as large as possible, so to enhance sorption performances, while significantly reducing dust emission. Commercial sintered getters are made, e.g., from zirconium (St 171®), from zirconium and St 707® (St 172), as well as titanium and St 101® (St 121) or titanium and St 707® (St 122). For mounting, either these getters are sintered onto metal foil as well carriers. Heating elements are commonly embedded and sintered into the material: for example, in the case of getter pump, sintered disks are stacked around a central heater which suitable for the activation needs.

## 2   Design of NEG Pumps

Ready to be installed combinations of getters and resistance heaters for activation are referred to as NEG pumps. Generally, the heating element is mounted in the center of a CF flange equipped with lead-through. Replaceable getter cartridges are fixed to the flange. The getter cartridges are built up either of folded getter strips or sintered getter discs (see Fig. 6). In both cases, the cartridges are designed to provide optimal access and contact between the getter surface and the gases to be bound. Getter strips, for example, are arranged in the cartridge as bellow-type folded rings similar to automobile air filters. The getter cartridge is a cylinder with a central hole for a heating element.

Special electrical power supplies (NEG pump controllers) are used for activation and are wire-connected to the pump via socket connectors. Larger getter pumps are equipped with temperature probes that allow automatic control of the getter material's temperature. One manufacturer's series (MK5, see Fig. 6 and Table 2) features a flange that is bakeable to $400°C$ after the socket connector is removed.

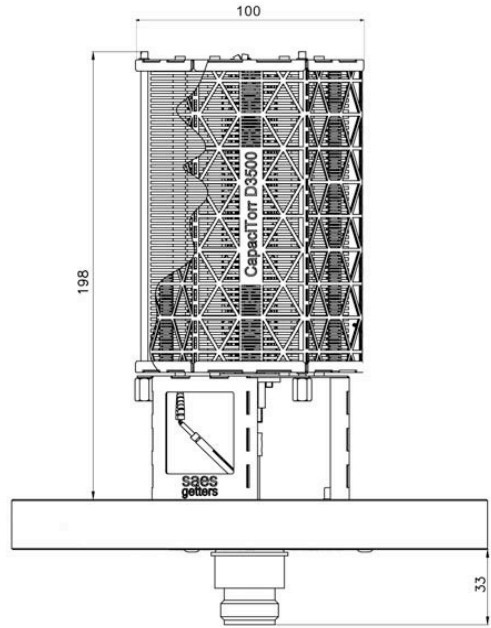

**Fig. 6:** Side view with partial section of a bulk getter pump (CapaciTorr D 3500 MK5 type). Courtesy of *SAES Getters S.p.A.*

**Table 2:** Nominal pumping speed $S_n$ for bulk getter pumps (*SAES* getters) of the design shown in Figs. 11 and 12; mounted nude in a vacuum system. More pumping speeds are available for the same flange, depending on the cartridge length. Getter material: St 172 sintered disks.

| Size | CF35 | CF63 | CF100 | CF150 |
|---|---|---|---|---|
| $S_n$ for $H_2$ in $\ell/s$ | 50/100/200/400 | 1000 | 1000/2000 | 2000/3600 |
| $S_n$ for CO in $\ell/s$ | 25/50/100/200 | 500 | 500/1000 | 1000/1800 |

The sorption curves for the largest NEG model for $H_2$ and CO are given in Fig. 7.

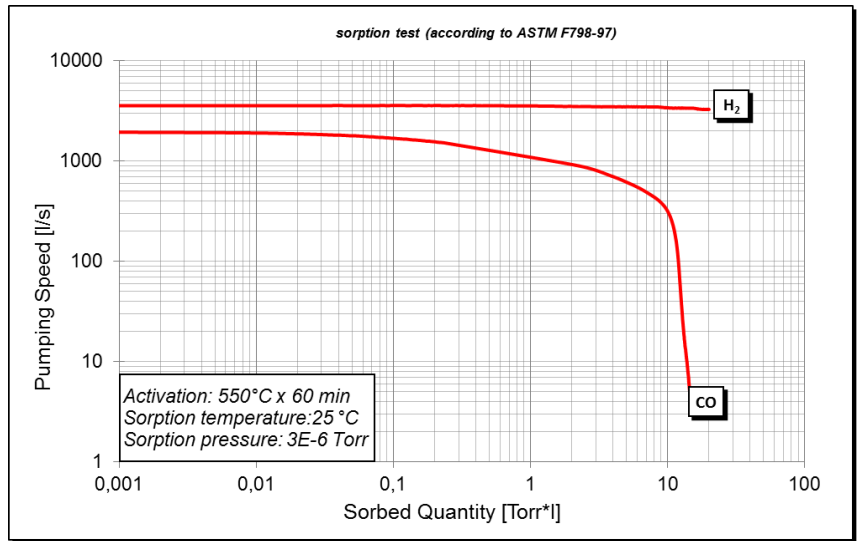

**Fig. 7:** Pumping speed vs sorbed amount for the CapaciTorr D 3500 NEG pump. Test gases are hydrogen and CO. Courtesy of *SAES Getters S.p.A.*

NEG pumps with much larger speeds can be produced for specific applications by arranging the getter elements in custom-designs [4-5].

NEG-alloys may also directly be applied to the inner vacuum chamber wall by coating [6]. By this method, which was developed and patented by CERN, the vacuum wall is turned from an outgassing source into a pump. It is helpful in particular where for geometrical reasons conductance to local pumps are small, e.g. for narrow insertion devices in high energy accelerators. The coating also reduces the outgassing rate and the secondary electron yield when the walls are exposed to photons (X-ray), electrons and ions bombardment. The getter is activated generally at 250°C x 2hours. However lower activation temperature (e.g. 180°C x 24 hours) are possible and can be applied especially for aluminum chambers which can undergo mechanical stresses and deformation when vacuum baked at temperature higher than 160°C. The NEG-alloy coating of the walls with a depth of about 1 μm is carried out by magnetron sputtering [6]. As the getter thickness is limited, the number of re-activation after venting is much less than for the NEG pumps. Also, pumping speed drops faster with multiple venting and is quoted to decrease as

$$S = 1/N \quad (S = \text{pumping speed}, N = \text{number of venting}). \qquad (3)$$

NEG coating is therefore an option in UHV-XHV systems which do not require frequent venting or reactivation.

## 3  NEG pumps, Pumping Speed and Getter Capacity

As described in Section 1, a bulk getter initially must undergo activation in vacuum and at high temperature before it shows any pumping action.

Pumping speed and getter capacity of volume getter pumps are usually measured at room temperature, a pressure of the gas to be bound of several $10^{-4}$ Pa ($10^{-6}$ mbar), and with free placement of the pump in a vacuum chamber (according for example to ASTM F798-82). Characteristics are exemplified using two gases with different binding behaviors: hydrogen and carbon monoxide. Pumping speed is measured in liters per second, getter capacity is usually determined as $\text{mbar}\ell$ ($pV$-value at $23°C$, see Section 2). For hydrogen that diffuses into the bulk of the getter material, pumping speed is nearly independent of the amount of gas already bound in the getter. For carbon monoxide and other gases that bind to the surface of the getter, pumping speed drops as reaction products gradually occupy more and more surface area. Thus, a getter can be characterized by giving the pumping speed with respect to the amount of previously sorbed gas. The amount of gas accumulated at the time when pumping speed has dropped to 10% or 20% of its initial value is termed getter capacity (see Fig. 5). In fact, differently from simple getters in sealed off devices, in the case of a NEG pump it is important that the pump can deliver a significant fraction of the nominal pumping speed. When the initial pumping speed drops below this minimum value, the pump needs to be re-activated.

Increasing temperature has an effect on all phases of gas binding. It promotes dissociation of gas molecules, and thus, a slight increase in pumping speed is generally observed at higher temperature.

As the gas content gradually approaches getter capacity, pumping speed of the getter for active gases drops accordingly. Depending on gas species, several reasons for this can be identified:

- The surface of the getter becomes covered with reaction products of getter material and absorbed gas, i.e., the surface passivates. Additional chemically active gas cannot react with the surface. Access of hydrogen that could dissolve in the getter is aggravated as well.
- The surface is clean but the concentration of hydrogen in the getter has grown too far (close to equilibrium concentration).

For both cases, reactivating is carried out by increasing the temperature:

- In reactivation (e.g., at $450°C$ for St 707®, $750°C$ for St 101®), chemically bound gases diffuse into the bulk of the getter material. Up to 20 reactivation cycles are possible until a noteworthy increase in concentration of bound gas in the bulk of the getter material starts to reduce pumping efficiency.
- Very high temperatures alter the equilibrium of hydrogen bound in the getter and surrounding hydrogen in the gas volume. Hydrogen partial pressure in the vacuum increases considerably and the hydrogen can be pumped down using mechanical pumps. After cooling, hydrogen can be absorbed anew. Theoretically, this procedure can be repeated indefinitely. However, small amounts of other gas species usually exist, which react at the surface.
- The third approach is termed continuous reactivation, i.e., after activation, the getter pump continues to operate permanently at higher temperature. This operating mode is suggested only for high gas loads (not $H_2$). Operating temperatures are well above activating temperatures (e.g., $280°C$ for St 707®, $400°C$ for St 101®).

## 4  Applications of NEG Pumps

NEG materials are used in many applications that require sealed vacua with long service life. These include any type of electron tubes [7], lamps, and stainless steel thermos flasks. Additionally, such materials are used for producing gas purifiers that purify process gases for semiconductor fabrication down to the ppt range [8]. NEG pumps have been so far mainly used in UHV applications because of their high pumping efficiency for hydrogen that limits the ultimate pressure in such applications and their compact size.

Applications of NEG pumps cover a variety of UHV laboratory systems, such as surface analysis equipment, photocathodes preparation and transportation, mass spectrometers as well as very large physical experiments [9-15]. NEG pumps are in fact extensively used in accelerators complex like synchrotron light sources, colliders or Free Electron Laser. In the industrial field NEG pumps are particularly helpful in electron gun based systems for scanning and transmission electron microscopes (SEM/TEM) or for electron lithography equipment for semiconductor processing or wafer inspection. where they provide larger pumping speed and a more compact package than sputter ion pumps.

As power consumption of getter pumps is low (after initial activation, further energy supply is generally unnecessary), they are frequently used in portable analysis equipment.

Bulk getter pumps can also be used to as a hydrogen source to adjust a particular hydrogen partial pressure in a vacuum by setting getter temperature.

As NEG pumps do not pump noble gases they are generally combined with other pumping devices, like cryo, turbo-molecular or sputter ion pumps. NEG elements and getter ion pumps have been traditionally combined together. The addition of a NEG cartridge or a NEG module inside the ion pump is a standard choice which improves the pumping speed for hydrogen and the other gases. However, in this configuration the total volume of the combined pump is still relatively large, due to the presence of the ion pump.

Novel designs have been proposed where the NEG cartridge acts as the main vacuum pump, supported by a small ion pump for the sorption of inert gases, which cannot be sorbed by the getter [16-18]. One of this design is showed in Fig. 8 [18].

Depending on the mounting flange, the pumping speed of the NEG cartridges varies from 100 to 2000 l/s ($H_2$) while the speed of the ion pump for argon is a few l/s. Such a speed is generally adequate as ungetterable gases (Argon and/or $CH_4$) are generally less than 1% of the total pressure of a UHV systems and do not require large pumping speed to be removed.

This approach is in line with the current trend in equipment miniaturization which calls for the reduction of the size of the vacuum components, pumps included. These combined pumps find adoption in all traditional NEG pump fields like accelerators, transportation suitcases, SEM/TEM, analytical equipment, where they provide a compact-low weight option at comparable high pumping speed.

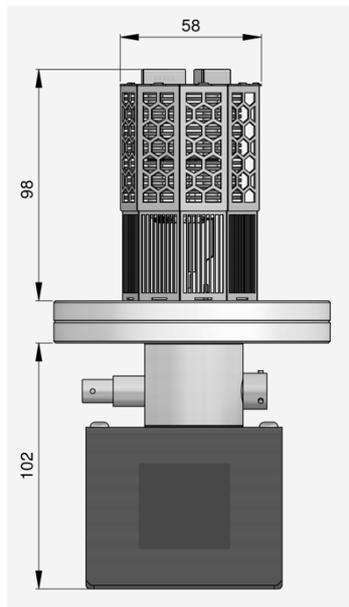

**Fig. 8:** Compact combination pump with a NEG element (500 l/S $H_2$) acting as the main pump for active gases, backed by a small ion pump (5 l/s $N_2$) for pumping of inert gases (NEXTorr D500-5). Courtesy of *SAES Getters S.p.A.*

## 5 Safety and Operating Recommendations

### 5.2 Air inrush during activation

During activation or reactivation of getter material, sudden air inrush must be prevented. At temperatures above 450°C (for St 101®) and 200°C (for St 707®), the getter material would react completely with the atmospheric oxygen. This combustion process is gradual but sustained in presence of an abundant source of oxygen or oxygenated species and would like to the complete oxidation of the getter material. Incomplete reaction is generally observed at lower temperature.

However, formation of a thick passivation layer may be observed in these cases, with a partial reduction of the getter capacity. In such a case, higher than standard activation temperatures are required to restore the getter pumping speed.

## 5.2   Venting

To avoid thermal runaway, it is safe to vent getter pump with reactive gases (air) at temperatures below $50°C$. Afterwards, the getter simply requires reactivation. In each venting, however, the getter sorbs gases irreversibly, and thus, its capacity is lower after reactivating. Use of nitrogen for venting is beneficial because pumping speed then remains high even after several reactivating cycles (after approximately 30 cycles of $N_2$ venting, $S$ still reaches 80% of $S_{initial}$; when using air for venting, it drops to 40 per cent of $S_{initial}$, see Fig. 9).

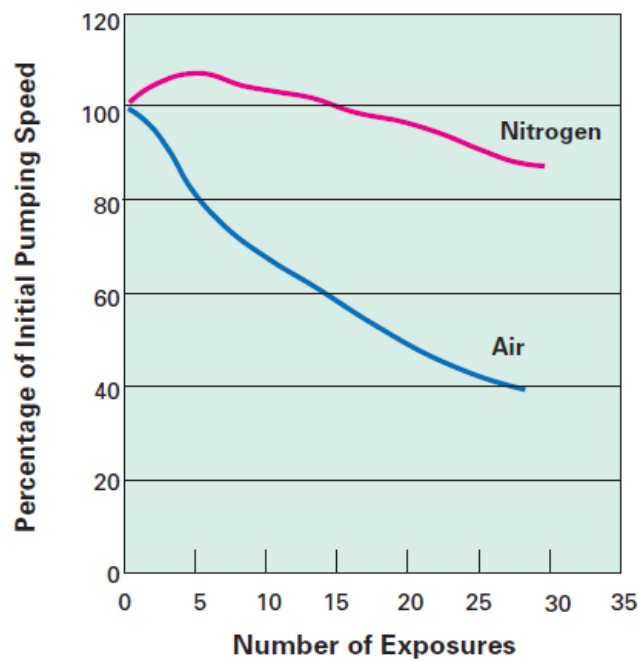

**Fig. 9:** Influence of the number of exposures $n_{exp}$ to air and nitrogen with subsequent reactivating on surface-related pumping speed $S/A$ for hydrogen. Getter material St 707® [1].

A further option would be venting with argon as protective gas. However, this procedure is rarely utilized in practice.

## 5.3   Embrittlement of getter material

Beyond certain hydrogen concentrations, hydrides form in the bulk getter material. This changes the crystalline structure and introduces lattice strain causing embrittlement of the material and ultimately the generation of cracks and structural damages of the getter disks as well as dust and fine powder. For traditional getter alloy, like St 101®, St 707® or St 172®, a safe limit for the embrittlement is generally set at 1500 Pa·l/g (≈10 Torr·l/g). Higher concentration levels may be tolerated in specific application or under defined working conditions. Should the getter pump need to absorb larger concentration of hydrogen different alloys have to be used. With this respect the ZAO®1, has an embrittlement limit 3 times larger and can be used safely up to 4500 Pa·l/g.

## 5.4 Trends in NEG pump technology

Differently from other more consolidated pumping technologies, NEG pumps have been subjected to a significant innovation process in the recent years. Innovations have addressed some of the critical issues of NEG technologies with the aim to either improve their performances or to extend their usability in a broader pressure range. New designs are now available in the following areas:
- NEG pump for High Vacuum ($10^{-8}$ mbar)
- Dust free NEG pumps.

### *5.4.1 NEG pumps for High Vacuum*

NEG pumps have been so far traditionally used in UHV. In fact, the use of conventional NEG pumps in high vacuum ($10^{-8}$-$10^{-7}$ mbar) would require too frequent reactivations (e.g. weekly) which might not be desired in several applications.

Increasing the getter working temperature to higher temperature, such as 200°C, is not recommended for commercial getters (e.g. St 172, St 707®) as the hydrogen equilibrium isotherm is close to $10^{-8}$ mbar (see for example Fig. 4). At 200°C the getter would therefore sorb active gases releasing however hydrogen and thus failing to keep the vacuum conditions.

Getter alloys with a much lower hydrogen equilibrium pressure are therefore required to solve this issue. Figure 10 compares St 707 equilibrium isotherms with those of the newly developed ZAO®1. This alloy has hydrogen equilibrium pressure in the $10^{-10}$ mbar range, fully compatible with the use in HV applications.

Figure 11 shows multiple sorption and reactivations cycles of a ZAO®1 NEG pump exposed to $1,3 \times 10^{-6}$ mbar of $CO_2$. Each sorption curves correspond to approximately one year continuous operation at $3 \times 10^{-8}$ mbar $CO_2$. After each sorption cycle the getter has been reactivated and a new sorption cycle started. Data show that the cycles are reproducible indicating the ability of the getter to effectively sink the gas load. The operational pressure range of NEG pumps based on ZAO®1 can therefore be extended from UHV to HV. The key features of NEG pumps in term of compactness, large pumping speed, ease of use can therefore be extended at least partially into the HV regime.

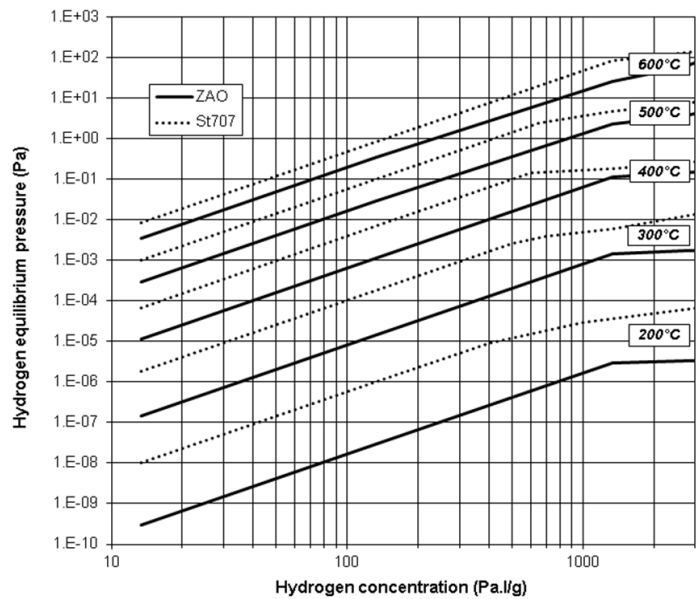

**Fig. 10:** Hydrogen equilibrium isotherms for St 707® and ZAO®1 alloys.

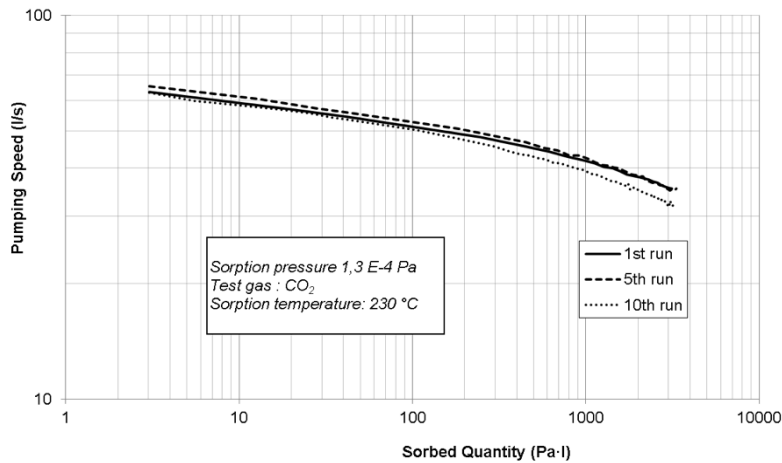

**Fig. 11:** Multiple sorption cycles of a ZAO®1 based pumps at moderate vacuum.

### 5.4.2 Dust free NEG pumps

NEG pumps using compressed getter powder are generally not recommended for particle sensitive applications. In presence of high electromagnetic fields, particles can in fact electrostatically charge and interfere, or can move and create shorts or even interact with the electron/ion beams and being vaporized. In semiconductor applications, particles can introduce defects on the silicon wafer and reduce the manufacturing yield. From this point of view sintered getters are significantly superior and are therefore used in electron microscopes, X-ray devices, accelerators and other applications where dust might be detrimental. In spite of this, in some cases, the cleanliness requirement of the application is such that even sintered St 172 disks are not used. A recent further improvement in this direction comes from the development of ZAO®1 sintered disks. Experimental tests carried out in several facilities [19], have showed that the dust emission from ZAO® pumps is negligible and it is compatible with very demanding applications, like for example mirror and monochromator chambers, superconductive radiofrequency cavities or inside high efficiency photocathode guns.